\documentclass[aps,preprint,superscriptaddress,prl]{revtex4}

\usepackage{graphicx}
\usepackage{amsmath}
\usepackage{amssymb}
\usepackage{units}
\usepackage{times}

\renewcommand{\vec}[1]{\boldsymbol{#1}}

\begin{document}

%
%

\title{Silicon surface with giant spin-splitting}

\author{I.~Gierz}
\affiliation{Max-Planck-Institut f\"ur Festk\"orperforschung,
D-70569 Stuttgart, Germany}
\author{T.~Suzuki}
\affiliation{Max-Planck-Institut f\"ur Festk\"orperforschung,
D-70569 Stuttgart, Germany}
\author{E.~Frantzeskakis}
\affiliation{Institut de Physique de la Mati{\`e}re Condens{\'e}e,
Ecole Polytechnique F{\'e}d{\'e}rale de Lausanne, CH-1015
Lausanne, Switzerland}
\author{S.~Pons}
\affiliation{Institut de Physique de la Mati{\`e}re Condens{\'e}e,
Ecole Polytechnique F{\'e}d{\'e}rale de Lausanne, CH-1015
Lausanne, Switzerland}\affiliation{D{\'e}partement Physique de la
Mati{\`e}re et des Mat{\'e}riaux, Institut Jean Lamour, CNRS,
Nancy Universit{\'e}, F-54506 Vandoeuvre-les-Nancy, France}
\author{S.~Ostanin}
\affiliation{Max-Planck-Institut f\"ur Mikrostrukturphysik,
D-06120 Halle (Saale), Germany}
\author{A.~Ernst}
\affiliation{Max-Planck-Institut f\"ur Mikrostrukturphysik,
D-06120 Halle (Saale), Germany}
\author{J.~Henk}
\affiliation{Max-Planck-Institut f\"ur Mikrostrukturphysik,
D-06120 Halle (Saale), Germany}
\author{M.~Grioni}
\affiliation{Institut de Physique de la Mati{\`e}re Condens{\'e}e,
Ecole Polytechnique F{\'e}d{\'e}rale de Lausanne, CH-1015
Lausanne, Switzerland}
\author{K.~Kern}
\affiliation{Max-Planck-Institut f\"ur Festk\"orperforschung,
D-70569 Stuttgart, Germany}\affiliation{Institut de Physique de la
Mati{\`e}re Condens{\'e}e, Ecole Polytechnique F{\'e}d{\'e}rale de
Lausanne, CH-1015 Lausanne, Switzerland}
\author{C.~R.~Ast}
\affiliation{Max-Planck-Institut f\"ur Festk\"orperforschung,
D-70569 Stuttgart, Germany}

\date{\today}

\begin{abstract}
We demonstrate the induction of a giant Rashba-type spin-splitting
on a semiconducting substrate by means of a Bi trimer adlayer on a
Si(111) wafer. The in-plane inversion symmetry is broken so that
the in-plane potential gradient induces a giant spin-splitting
with a Rashba energy of about 140\,meV, which is more than an
order of magnitude larger than what has previously been reported
for any semiconductor heterostructure. The separation of the
electronic states is larger than their lifetime broadening, which
has been directly observed with angular resolved photoemission
spectroscopy. The experimental results are confirmed by
relativistic first-principles calculations. We envision important
implications for basic phenomena as well as for the semiconductor
based technology.
\end{abstract}

\maketitle

%
%

Exploiting the electron spin for information processing is one of
the leading goals in the rapidly growing field of spintronics. At
its heart lies the Rashba-Bychkov (RB) type spin-splitting, where
the spin-orbit interaction lifts the spin degeneracy in a symmetry
broken environment \cite{Rashba}. Many device proposals make use
of this concept \cite{Sinova,Datta,Ohe,Koga} with some interesting
proofs of principle \cite{Kato,Wolf}. The materials of choice are
semiconductor heterostructures, albeit the size of the
spin-splitting is typically very small. A large spin-splitting is
desirable as it would, for example, decrease the precession time
of the spin in a spin transistor \cite{Datta} so that it is
smaller than the spin relaxation time. Furthermore, a separation
of the spin-split states beyond their lifetime broadening is an
important criterion for distinguishing between the intrinsic and
extrinsic spin Hall effect \cite{Sinova,Wunderlich,Bernevig}. The
different influences on the intrinsic spin Hall conductivity, such
as disorder and elastic/inelastic lifetime, are still under debate
\cite{Wang,Yang}.

Recently, a giant spin-splitting has been demonstrated for noble
metal based surface alloys \cite{Ast,Ast1,Ast2,Pacile}, where
heavy elements with a strong atomic spin-orbit coupling are
incorporated into the surface. These systems, however, are not
suitable for the field of spintronics because of the presence of
spin-degenerate bands at the Fermi level originating from the
metallic substrate. One possible alternative is to grow thin films
with spin-split bands onto a semiconducting substrate
\cite{Frantzeskakis,He,Hirahara}. However, due to confinement
effects a multitude of quantum well states arise, which
potentially influence the transport properties of the system. It
is, therefore, desirable to transfer the concept of the giant
spin-splitting directly onto a semiconductor surface.

Here we show that a monolayer of Bi trimers on a Si(111) surface
forms a two-dimensional (2D) electronic structure with a giant
spin-splitting much larger than what has been observed so far at
the interfaces of semiconductor heterostructures. The effect can
be traced to a strong contribution of an in-plane potential
gradient due to an inherent structural inversion asymmetry (RB
model). While the structure of this system has been studied both
theoretically as well as experimentally \cite{Miwa,Wan,Shioda},
the electronic structure, in particular a possible spin-splitting
of the electronic states, has remained a controversial issue
\cite{Kinoshita,Kim1}. We demonstrate unequivocally that Bi
induces a giant spin-splitting at the silicon surface.
Furthermore, the spin-splitting is observed to be larger than the
lifetime broadening, so that the Bi/Si(111) system is a prime
candidate for spintronics applications or studying the intrinsic
spin Hall effect. In addition, the silicon substrate allows for
excellent compatibility with existing silicon-based semiconductor
electronics.

\begin{figure}
  \includegraphics[width = 0.5\columnwidth]{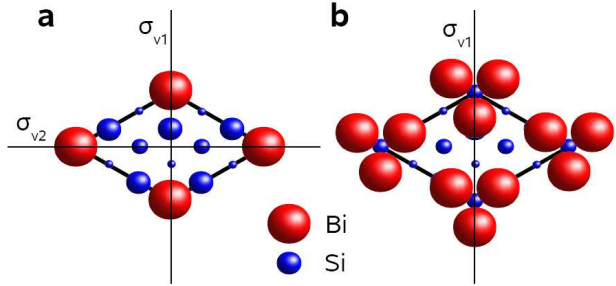}
  \caption{(color online) Structural model of the two ($\sqrt3 \times\sqrt3$)R30$^\circ$
  phases of Bi/Si(111): (a) monomer phase (b) trimer phase. The thin
  black lines indicate mirror planes of the Bi adlayer. The thicker black lines
  indicate the ($\sqrt3 \times\sqrt3$)R30$^\circ$ unit cell. The smaller
  the spheres, the further away they are from the surface.}
  \label{fig:structure}
\end{figure}

A single layer of Bi on Si(111) grows in a monomer as well as a
trimer configuration, both of which show a ($\sqrt 3\times\sqrt
3$)R30$^\circ$ reconstruction \cite{Miwa,Wan,Shioda}. A structural
model is shown in Fig.\ \ref{fig:structure} for the monomer phase
(a) and the trimer phase (b). Both the monomers and the trimers
are centered on top of second layer Si atoms ($T_4$ lattice sites)
forming a symmorphic space group based on the point group $3m$.
The Si substrate breaks the in-plane inversion symmetry for both
the monomer and the trimer phase. Looking at the isolated Bi
adlayer alone, the trimer formation introduces a reduction of the
symmetry because the mirror plane $\sigma_{v2}$ is missing. The
mirror plane $\sigma_{v1}$ holds for both the monomer and the
trimer phase as well as for the combination of adlayer and Si
substrate. From these simple symmetry considerations we conclude
that the Bi trimer phase is the least symmetric structure and,
hence, should lead to the bigger spin-splitting. We, therefore,
only consider the trimer phase. Its preparation was verified with
quantitative low-energy electron diffraction measurements
\cite{EPAPS}.

\begin{figure}
  \includegraphics[width = 0.5\columnwidth]{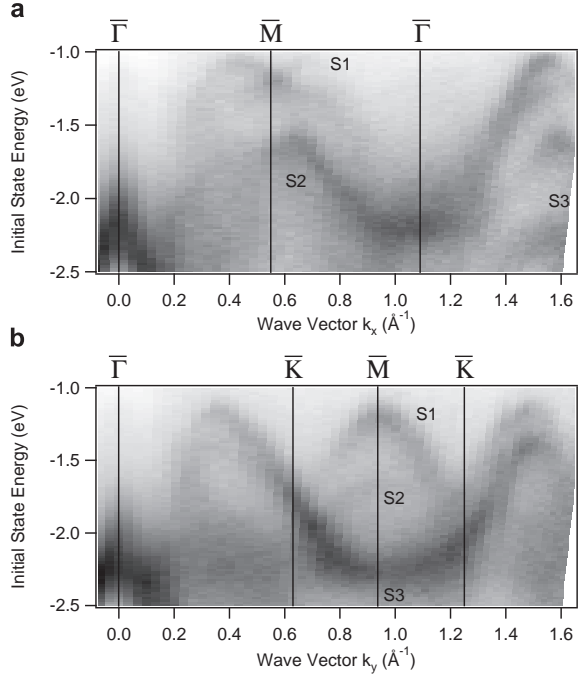}
  \caption{The two panels show angle-resolved ultra violet photoemission
  spectroscopy data of the trimer phase of bismuth on Si(111) along the
  two high symmetry directions $\overline{\Gamma}\,\overline{\mbox{M}}$
  The photoemission intensity is on a linear scale with black and
  white corresponding to highest and lowest intensity, respectively.
  The energy scale is set to zero at the Fermi level. (a)
  and $\overline{\Gamma}\,\overline{\mbox{K}}\,\overline{\mbox{M}}$ (b). Light an dark
  areas correspond to low and high intensities, respectively. A splitting of
  the two-dimensional state into two bands around the $\overline{\mbox{M}}$ point along
  the $\overline{\Gamma}\,\overline{\mbox{M}}$-direction at an initial state
  energy of about $-1.3\,$eV is clearly visible in
  panel (a). We attribute this splitting to the Rashba-Bychkov
  effect with a momentum offset $k_0= 0.126$\,\AA $^{-1}$
  and a Rashba energy $E_R= 140$\,meV.}
  \label{fig:experiment}
\end{figure}

The experimental band structure measured with angular resolved
photoemission spectroscopy (ARPES) along the two high symmetry
directions of the surface Brillouin zone (SBZ)
$\overline{\Gamma}\,\overline{\mbox{M}}$ and
$\overline{\Gamma}\,\overline{\mbox{K}}\,\overline{\mbox{M}}$ is
displayed in Fig.\ \ref{fig:experiment} (a) and (b), respectively
\cite{EPAPS}. The intense feature near $\overline{\Gamma}$ at an
energy of about $-2.3$\,eV can be attributed to the silicon bulk.
The other features (S1, S2, S3) in Fig.\ \ref{fig:experiment}(a)
originate from the 2D electronic structure of the surface. S1 is
most intense at the $\overline{\mbox{M}}$-point at an initial
state energy of about $-1.3$\,eV. This band splits in two
components when moving away from the high symmetry point
$\overline{\mbox{M}}$, which is a strong indication of a RB-type
spin-splitting. S2 is located at about $-2.3$\,eV at the second
$\overline{\Gamma}$-point and disperses upwards towards the
$\overline{\mbox{M}}$-points. The third state S3 shows the highest
intensity at the second $\overline{\mbox{M}}$-point at an energy
of about $-2.5$\,eV. This band moves downwards in energy towards
the second $\overline{\Gamma}$-point. The bandwidth of S3 is
smaller than the one for S2. These three 2D states are also
visible along the
$\overline{\Gamma}\,\overline{\mbox{K}}\,\overline{\mbox{M}}$-direction
as shown in Fig. \ref{fig:experiment} (b). S1 appears as a
parabolic band with negative effective mass with a band maximum
located at about $-1.3$\,eV at the $\overline{\mbox{M}}$-point.
Along the
$\overline{\Gamma}\,\overline{\mbox{K}}\,\overline{\mbox{M}}$
direction no splitting of this band has been observed. S2 is
located around $-1.8$\,eV at $\overline{\mbox{M}}$, but only with
a very weak intensity. The most intense feature along the
$\overline{\Gamma}\,\overline{\mbox{K}}\,\overline{\mbox{M}}$-direction
is the S3 2D state with a band minimum at about $-2.5$\,eV at the
$\overline{\mbox{M}}$-point and an upwards dispersion towards the
neighboring $\overline{\mbox{K}}$-points.

\begin{figure}
  \includegraphics[width = 0.45\columnwidth]{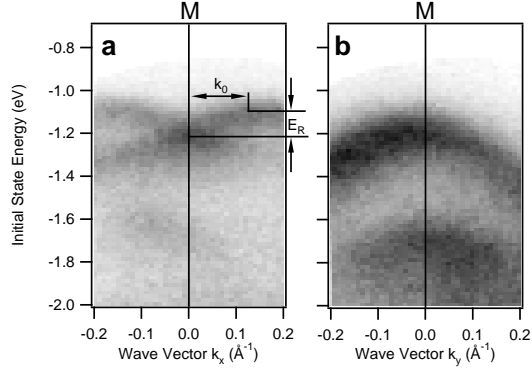}
  \caption{Experimental band structure of Bi/Si(111) near the
  $\overline{\mbox{M}}$-point. The measurements along
  $\overline{\Gamma}\,\overline{\mbox{M}}\,\overline{\Gamma}$ (a) and
  $\overline{\mbox{K}}\,\overline{\mbox{M}}\,\overline{\mbox{K}}$ (b) show the
  anisotropic topology of the spin-split bands.}
  \label{fig:CloseUp}
\end{figure}


A possible spin-splitting in the Bi/Si(111) system is an
unresolved issue in the literature. While Kinoshita {\it et al.}
\cite{Kinoshita} consider a splitting in the three 2D states
related to a strong spin-orbit interaction of the Bi atoms, it has
been dismissed by Kim {\it et al.} \cite{Kim1}. In the following,
we will show from the experimental data as well as spin-resolved
band structure calculations that the band structure shows a giant
spin-splitting of the electronic states due to the RB effect.

Spin-degeneracy is a consequence of both time reversal and spatial
inversion symmetry. If the latter is broken spin-degeneracy can be
lifted by the spin-orbit interaction (RB model) \cite{EPAPS}. In
addition, if the inversion symmetry is also broken in the plane of
the two-dimensional electron gas, the contribution from an
in-plane gradient can strongly enhance the spin-splitting
\cite{Ast1}. The characteristic parameters quantifying the
strength of the splitting are the momentum offset $k_0$, the
coupling constant in the Hamiltonian $\alpha_R$ (Rashba
parameter), as well as the Rashba energy $E_R$. They are related
by $E_R={\hbar^2 k_0^2}/{2m^*}$ and $k_0=m^*\alpha_R/\hbar^2$.
Here $m^*$ is the effective mass.

A close up of the band structure near the
$\overline{\mbox{M}}$-point is shown in Fig.\ \ref{fig:CloseUp}.
The bands along $\overline{\Gamma}\,\overline{\mbox{M}}$ (Fig.\
\ref{fig:CloseUp}(a)) near $-$1.2\,eV clearly show the
characteristic dispersion of a RB type spin-splitting with the
band crossing at the $\overline{\mbox{M}}$-point and the shift of
the maxima away from it. From the data we extract the momentum
offset $k_0=0.126$\,\AA$^{-1}$, an effective mass of
$m^*=0.7\,m_e$ ($m_e$ free electron mass) as well as the Rashba
energy $E_R=140$\,meV. From these values we can calculate the
Rashba parameter $\alpha_R$=1.37\,eV\AA. The spin-splitting is
well resolved in the data. The average line width for the
spin-split states at the band maximum ($k_x=-0.126\,$\AA$^{-1}$)
is 195\,meV, which accounts for intrinsic lifetime as well as
interactions and scattering. The separation of the states is about
220\,meV.

The spin-splitting at the $\overline{\mbox{M}}$-point in Fig.\
\ref{fig:CloseUp}(a) is strongest along the
$\overline{\Gamma}\,\overline{\mbox{M}}$-direction. Along the
$\overline{\mbox{K}}\,\overline{\mbox{M}}\,\overline{\mbox{K}}$-direction
in Fig.\ \ref{fig:CloseUp}(b) the spin-splitting at the
$\overline{\mbox{M}}$-point is much weaker and cannot be resolved
in the experiment. This peculiar band topology can be related to
the symmetry properties of the $\overline{\mbox{M}}$-point. As the
$\overline{\mbox{M}}$-point is located on the border of the first
SBZ it has a reduced symmetry as compared to the
$\overline{\Gamma}$-point. Despite the symmetry breaking of the Bi
trimers, the mirror symmetry $\sigma_{v1}$ (see Fig.\
\ref{fig:structure}) holds so that for the dispersion along the
$\overline{\mbox{K}}\,\overline{\mbox{M}}\,\overline{\mbox{K}}$-direction
the spin-splitting is greatly reduced, i.\ e.\ it can not be
observed in the data.

\begin{figure}
  \includegraphics[width = 0.5\columnwidth]{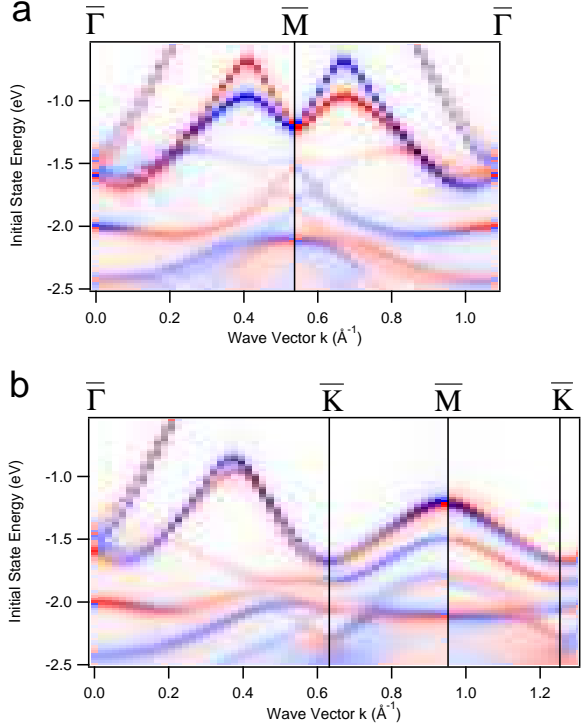}
  \caption{(color online) Theoretical band structure calculations for the trimer
  phase of bismuth on silicon(111). Panel (a) and (b) show the calculated
  dispersion along $\overline{\Gamma}\,\overline{\mbox{M}}$ and
  $\overline{\Gamma}\,\overline{\mbox{K}}\,\overline{\mbox{M}}$, respectively. Blue
  and red correspond two opposite spin-polarizations. The calculated spectra
  reproduce the main features of the measured band structure, especially the
  spin-splitting of the bands around the $\overline{\mbox{M}}$-point along
  $\overline{\Gamma}\,\overline{\mbox{M}}$.}
  \label{fig:calculation}
\end{figure}

To support our interpretation of the observed spin-splitting, we
conducted spin-resolved first principles band structure
calculations, which were performed in close analogy to our
previous calculations on the RB effect \cite{Ast1}. The surface
geometry of the trimer structure is determined from first
principles using the \textsc{Vienna Ab-initio Simulation Package}
(VASP) which provides precise total energies and forces
\cite{Kresse96}. The Bi trimers (milkstool structure) are relaxed
outward by $\unit[13]{\%}$ from the ideal positions (100\%
corresponds to the Si bulk interlayer distances, lattice constant
$\unit[5.403]{\AA}$). The subsurface relaxations are small ($<
\unit[0.5]{\%}$) and neglected in the Korringa-Kohn-Rostoker (KKR)
calculations. The in-plane displacement of the Bi trimer atoms
$\delta$ is $0.3$, with $\delta = 0$ indicating Bi on-top of first
layer Si atoms and $\delta = 1$ coinciding Bi-trimer atoms on
$T_4$ sites. The subsequent KKR and relativistic layer-KKR
calculations use the structural data from VASP as input. The
spectral density $n_{\pm}(E, \vec{k}_{\parallel})$ is obtained
from the imaginary part of the site-dependent Green function.
Resolved with respect to spin orientation (index $\pm$) and
angular momentum, it allows a detailed analysis of the electronic
structure. The difference $n_{+}(E, \vec{k}_{\parallel}) -
n_{-}(E, \vec{k}_{\parallel})$ reveals the characteristic spin
splitting of RB-split bands.

The results of the band structure calculations are shown in Fig.\
\ref{fig:calculation} for the
$\overline{\Gamma}\,\overline{\mbox{M}}$-direction in (a) and for
the $\overline{\Gamma}\,\overline{\mbox{K}}$-direction in (b). The
intensity scale shows the total spectral density ($n_{+}(E,
\vec{k}_{\parallel})+n_{-}(E, \vec{k}_{\parallel})$) of the states
multiplied by the sign of the spin-polarization sgn($n_{+}(E,
\vec{k}_{\parallel})-n_{-}(E, \vec{k}_{\parallel})$) perpendicular
to the high symmetry line, i.\ e.\ blue and red colors correspond
to opposite spin-polarizations. The calculations reproduce all the
main features of the measured band structure. In particular, the
splitting of the S1 band around the $\overline{\mbox{M}}$-point
along the $\overline{\Gamma}\,\overline{\mbox{M}}$-direction is
well documented. As can be seen in figure \ref{fig:calculation}
the two branches of the split S1 band clearly show opposite
spin-polarization, i.~e. a giant spin-splitting in the electronic
structure of Bi/Si(111).

The spin-splitting is strongly anisotropic around
$\overline{\mathrm{M}}$. The peculiar band topology, which was
observed in the experiment is clearly reproduced in the
calculations. This can again be attributed to the lower symmetry
of wave vectors $\vec{k}_{\parallel}$ within
($\overline{\Gamma}\,\overline{\mathrm{M}}$) or perpendicular
($\overline{\Gamma}\,\overline{\mathrm{K}}\,\overline{\mathrm{M}}$)
to a mirror plane of the system. It is conceived that this feature
results from the `trimerization' of the three Bi sites in the 2D
unit cell; calculations with a reduced $\delta$ (i.~e. larger
distance between Bi trimer atoms) indicate an even smaller
splitting along
$\overline{\Gamma}\,\overline{\mathrm{K}}\,\overline{\mathrm{M}}$.
Furthermore, the calculations show that about 83\% of the
spin-split states at the $\overline{\mbox{M}}$-point are localized
in the Bi adlayer and about 16\% in the first Si layer. One can
thus speculate that the spin-splitting is particularly influenced
by the Bi adlayer and that trimerization symmetry breaking
increases the effect of the in-plane potential gradient.

The giant spin-splitting in the Bi/Si(111) trimer system has a
similar origin as in the Bi/Ag(111) surface alloy: An inversion
symmetry breaking in the plane of the surface leads to a strong
contribution from an in-plane potential gradient, which
substantially enhances the spin-splitting. In both systems the
threefold symmetry of the underlying substrate breaks the in-plane
inversion symmetry. However, considering only the topmost layer,
the trimer formation in Bi/Si(111) also leads to a breaking of the
in-plane inversion symmetry (see Fig.\ \ref{fig:structure}), which
is not the case for the Bi/Ag(111) surface alloy.

Comparing the spin-splitting of the Bi/Si(111) electronic
structure to semiconductor heterostructures, we find that in the
latter the spin-splitting is substantially smaller. For example,
for an inverted InGaAs/InAlAs heterostructure a Rashba constant of
$\alpha_R=0.07$\,eV\AA\ has been measured \cite{Nitta}. With an
effective mass of $m^*=0.05\,m_e$, a Rashba energy of
$E_R=16\,\mu$eV can be calculated. For HgTe quantum wells a Rashba
constant $\alpha_R=0.45$\,eV\AA\ has been found \cite{Schultz}.
However, here the spin-splitting has been identified to be
proportional to $k_{||}^3$ instead of a linear dependence
\cite{Zhang}. For the Bi/Si(111) system, the Rashba energy
$E_R=140$\,meV as well as the Rashba parameter
$\alpha_R=1.37\,$eV\AA\ are much bigger. From the momentum offset
$k_0=0.126\,$\AA$^{-1}$ we can calculate that a phase shift of the
spin precession angle $\Delta\theta=\pi$ can be obtained after a
length $L=\Delta\theta/2k_0$ of only $1.3$\,nm. In the
InGaAs/InAlAs heterostructure a length of 400\,nm has been
estimated. While these figures show the excellent potential of the
Bi/Si(111) system, additional measurements giving insight into the
transport properties, such as Shubnikov-de Haas oscillations, are
necessary to further elaborate the suitability of this system for
spintronics applications. Corresponding experiments are in
progress.

We have shown that the trimer phase of Bi on Si(111) shows a giant
spin-splitting. The experimental results reveal the characteristic
band dispersion of a RB-type spin-splitting with a peculiar band
topology at the $\overline{\mbox{M}}$ point. They are confirmed by
first principles band structure calculations. The splitting is
caused by the spin-orbit interaction induced RB effect in
combination with a strong contribution from the in-plane gradient
due to the reduced symmetry of the trimer structure and the
substrate. Furthermore, this spin-splitting is of the same order
of magnitude as the one reported for Bi/Ag(111) and orders of
magnitude lager than a typical spin-splitting reported for
semiconductor heterostructures. In this way, we have transferred
the concept of giant spin-splitting onto a semiconducting
substrate. This gives excellent perspectives for the use of this
concept in the field of spintronics. In particular, the silicon
substrate makes this system compatible with existing semiconductor
technology. On the fundamental side such systems are interesting
for, e.~g., the spin Hall effect. Since the energy separation of
the spin-split states (220\,meV) is larger than the lifetime
broadening (195\,meV), it may be easier to distinguish the
extrinsic and intrinsic spin Hall effects.

%
%

\section*{Supporting Information}

\subsection*{Sample preparation}

The sample preparation as well as the measurements were conducted
in ultra high vacuum (UHV) with a base pressure of $2\times
10^{-10}$\,mbar. The n-doped Si(111) substrate was annealed at
1100$^{\circ}$C by direct current heating for 10 minutes and
cooled down slowly to 800$^\circ$C over a time interval of 10
minutes until a sharp (7$\times$7) low energy electron diffraction
(LEED) pattern was observed. One monolayer (ML) of Bismuth was
deposited at a substrate temperature of 470$^\circ$C using a
commercial electron beam evaporator to form an adlayer of trimers.
The substrate temperature was measured with an optical pyrometer.
After Bi-deposition LEED measurements showed a ($\sqrt
3\times\sqrt 3$)R30$^\circ$ reconstruction.

\subsection*{Quantitative LEED measurements}

\begin{figure}
  \includegraphics[width = 0.5\columnwidth]{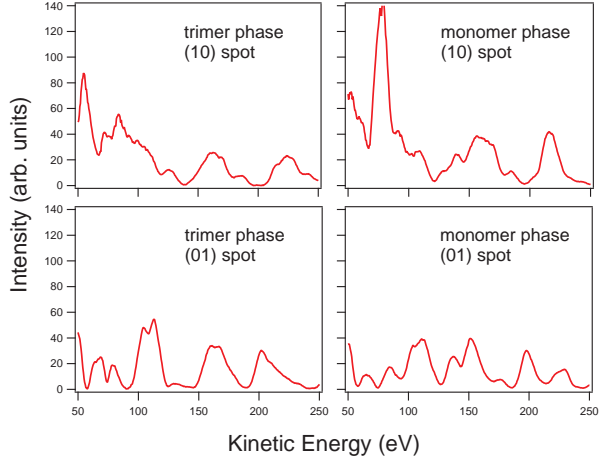}
  \caption{The bismuth coverage was determined via quantitative LEED
  measurements. The figures show the integrated intensity as a function
  of electron energy for the (10) and (01) spots of the monomer and
  trimer phase, respectively.}
  \label{fig:LEED}
\end{figure}

As both the monomer phase at 1/3 ML Bi coverage and the trimer
phase at 1 ML Bi coverage show the same ($\sqrt 3\times\sqrt
3$)R30$^\circ$ reconstruction we used quantitative LEED
measurements to distinguish between the two phases. We measured
the integrated intensity of the (10) and (01) spots as a function
of electron energy and compared them to calculations done by Wan
{\it et al.} \cite{Wan}. The measured data was averaged over three
equivalent spots and smoothed. The result is shown in figure
\ref{fig:LEED}. The agreement between measured and calculated
spectra is quite convincing, allowing an unambiguous
identification and preparation of the trimer phase.

\subsection*{STM measurements}

\begin{figure}
  \includegraphics[width = 0.5\columnwidth]{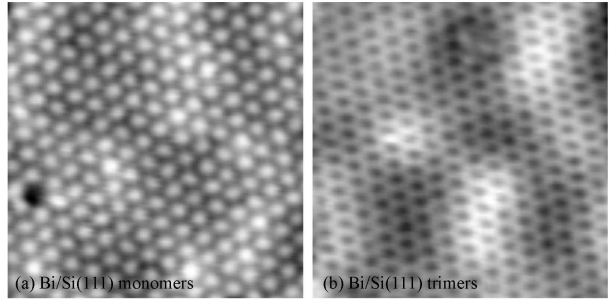}
  \caption{Topographic 10$\times$10\,nm$^2$ STM images of the
  ($\sqrt3 \times\sqrt3$)R30$^\circ$ phase of Bi/Si(111).
  The monomer phase and the trimer phase are shown in (a) and (b), respectively.}
  \label{fig:STM}
\end{figure}

In figure \ref{fig:STM} two (10$\times$10)nm$^2$ topographic STM
images are shown. In (a) the monomer phase is of the Bi/Si(111)
structure is shown at a bias voltage of $-$1.55\,V and a tunneling
current of 0.2\,nA is shown. In (b) the trimer phase is shown at a
bias voltage of $-$0.95\,V. The tunneling current was 0.2\,nA as
well.

\subsection*{Band structure measurements}

\begin{figure}
  \includegraphics[width = 0.5\columnwidth]{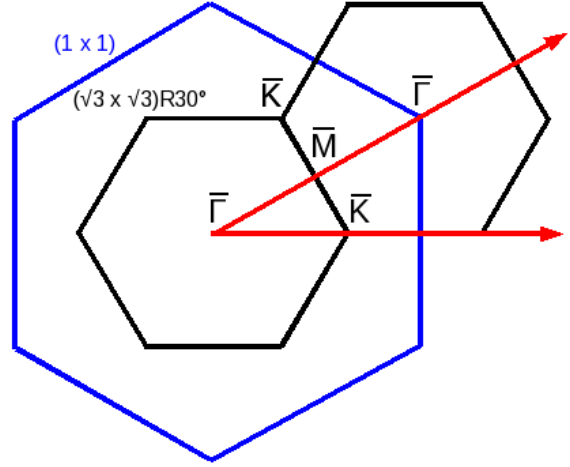}
  \caption{Sketch of the surface Brillouin zone (SBZ): the blue and
  black hexagons show the (1$\times$1) SBZ of the Si substrate and the
  ($\sqrt3\times\sqrt3$)R30$^\circ$ SBZ of the trimer phase of Bi/Si(111),
  respectively. The red arrows indicate the two high symmetry directions
  $\overline{\Gamma}\,\overline{\mbox{M}}$
  and $\overline{\Gamma}\,\overline{\mbox{K}}\,\overline{\mbox{M}}$
  with respect to the ($\sqrt3\times\sqrt3$)R30$^\circ$ SBZ.}
  \label{fig:SBZ}
\end{figure}

Angle-resolved photoemission spectroscopy (ARPES) measurements
were done with a hemispherical SPECS HSA3500 electron analyzer
with an energy resolution of $\sim$10meV. The images were recorded
with a step size of 1$^\circ$ in angular direction. We used
monochromatized He I radiation with an energy of 21.2\,eV for the
ARPES measurements. The sample was kept at 90\,K during the
measurements. We measured the band structure along the two high
symmetry directions $\overline{\Gamma}\,\overline{\mbox{M}}$ and
$\overline{\Gamma}\,\overline{\mbox{K}}\,\overline{\mbox{M}}$ of
the surface Brillouin zone (SBZ) associated with the
($\sqrt3\times\sqrt3$)R30$^\circ$ reconstruction (see Fig.\
\ref{fig:SBZ}).

The zero for the initial state energy in the measured data was set
to the Fermi level. We chose an n-doped Si substrate because the
position of the Fermi level is close to the conduction band. This
allows states within the bulk band gap to be easily accessible by
photoemission spectroscopy. From the recorded data we subtracted a
Shirley background and normalized every energy distribution curve
(EDC) of the image to the same integrated intensity.

\subsection*{Rashba-Bychkov model}

Spin-degeneracy is a consequence of both time reversal and spatial
inversion symmetry. If the latter is broken spin-degeneracy can be
lifted by the spin-orbit interaction. For a free electron gas in
two dimensions the spin-orbit coupling Hamiltonian in the
Rashba-Bychkov (RB) model is given by the following equation
\cite{Rashba}:
$$H_{SO}=\alpha_R\vec{\sigma}(\vec{k_{||}}\times \vec{e_z})$$
where the coupling constant $\alpha_R$ is called Rashba parameter,
$\vec{\sigma}$ are the Pauli matrices, $\vec{k_{||}}$ is the
in-plane momentum and $\vec{e_z}$ is a unit vector normal to the
plane of the two dimensional electron gas (2DEG). The resulting
energy dispersion is:
$$E(k_{||}) = \frac{\hbar^2}{2m^{*}}(k_{||}\pm k_0)^2+E_0$$
with the effective mass $m^{*}$, the offset in parallel momentum
$k_0$ and the band extremum at $E_0$. The RB Hamiltonian lifts the
spin-degeneracy and introduces an offset in parallel momentum of
the nearly free electron parabola. At high symmetry points the
bands are degenerate due to time inversion symmetry. As a result
the band dispersion of a RB split band has a very characteristic
shape. The Rashba parameter $\alpha_R$ in the most straight
forward formulation of the RB Hamiltonian is proportional to the
potential gradient perpendicular to the plane of the 2DEG. An
additional contribution from an in-plane gradient can strongly
enhance the spin-splitting.

From the photoemission data the Rashba energy $E_R$ (the energy
difference between the crossing point of the two parabolas and the
band extremum), the offset of the two parabolas in parallel
momentum $k_0$ and the effective mass $m^{*}$ can be determined.
These parameters are related to an effective Rashba-parameter
$\alpha_R$ via
\begin{equation}
E_R =\frac{\hbar^2 k_0^2}{2m^*}= \frac{1}{2}\alpha_R k_0
\label{equation}
\end{equation}
This effective Rashba parameter $\alpha_R$ contains all the
different contributions to the spin-splitting, i.~e. the potential
gradient in the plane of as well as perpendicular to the surface
and the atomic spin-orbit coupling from the atoms involved. The
Rashba parameter $\alpha_R$ determined from equation
(\ref{equation}) should be understood as a characteristic
parameter for comparison with other systems.

\end{document}